# Lagrangian description and Hamiltonian density for the electrodynamics of dispersive metamaterials

## Pi-Gang Luan

*Department of Optics and Photonics, National Central University, Jhongli District 32051, Taoyuan City, Taiwan, Republic of China*

**Abstract**

**The Lagrangians and dissipation functions are proposed for use in the electrodynamics of the double-negative and chiral metamaterials with finite loss. The double-negative metamaterial considered here is the 'wires and split-rings' periodic structure, while the chiral metamaterial is the 'single-resonance helical resonators' array. For either system, application of Legendre transformation leads to a Hamiltonian density identical to the energy density obtained in our previous work based on the Poynting theorem and the mechanism of the power loss. This coincidence implies the correctness of the energy density formulas obtained before. The Lagrangian description and Hamiltonian formulation can be further developed for exploring the properties of the elementary excitations or quasiparticles in dispersive metamaterials due to light-matter interactions.**

## 1. INTRODUCTION

A dispersive material is a material whose constituent atoms or molecules respond to the incident light dynamically, so that the collective effects such as refractive index and permittivity or permeability are frequency dependent. In addition to dispersion, real materials are inherently dissipative since energy can be carried away by other excitations such as scattered phonons. For the electromagnetic system itself, the dissipative power is expressed as the power loss [1]. Over the past two decades, artificial structures named metamaterials composed of 'artificial atoms' made of split rings or helical resonators have aroused great interest due to their potential utilities in the microwaves and photonics technologies. As effective media, metamaterials are also dispersive, and usually highly dissipative, so they can be categorized as dispersive and absorptive media [2,3].

An important problem about dispersive media is how to identify the electromagnetic energy density in them [1,4,5]. The solution of this problem has practical significance because it may provide relevant knowledge for nano-devices design [6]. It also has theoretical interest because the fundamental problem of energy propagation velocity related to causality and relativity is based on it [7-9]. A number of solutions for this problem have already been proposed [4,5,7-20], but so far most of them dealt only with the dispersion of permittivity, without considering the corresponding dispersion of the permeability [4,5,16-18,20]. Besides, some results are controversial [16]. For example, there is no consensus on whether a negative energy density is physically acceptable, although most believe energy density should be positive [4-15,18-20].

In our previous works we proposed a method for deriving the energy density based on the Poynting theorem and the knowledge of the power loss mechanism [14,15]. Our method not only can identify the energy density, but also have helped to resolve the problem of the contradictory results derived by the equivalent circuits (EC) [11,12] and the electrodynamics (ED) methods [13].

In this paper, we introduce the Lagrangian density [21-23] and dissipation function [24] density for the Lagrangian description of the electrodynamics in two kinds of typical dispersive and absorptive metamaterials. The first metmaterial we considered is the 'double-negative metamaterial' composed of wires [25] and split-rings [26] array, which can have negative refractive index in some frequency range when both the permittivity and permeability are negative [27-29]. The second one is the chiral metamaterial of 'single-resonance helical resonators array', which yields two refractive indices of opposite signs corresponding to the right-hand and left-hand circularly polarized propagating mode, respectively [30-32]. We will show in this paper that the Hamiltonian densities corresponding to these two metamaterials are identical to the energy densities derived in our previous works [14,15] except for the irrelevant divergence terms. The purpose of this study is twofold. The first one is to provide an alternative way for deriving the energy density. The second one is to construct the effective Lagrangian and Hamiltonian for further studying the low energy quasiparticles or elementary excitations in these media when quantization procedures are properly implemented.

## 2. THE DRUDE MODEL

As a warm-up example, we first try to construct the Lagrangian density and dissipation function density for the Drude model of the plasma system and discuss its Hamiltonian density. We will learn essential knowledge from this example and then apply some similar ideas to the study of the dispersive metamaterials. The electrons in Drude model are assumed to be non-interacting among themselves, accelerated by the applied electric field, and dissipate their momenta and energies through the collisions with phonons, dislocations, and defects in the ions background. The average effect of the dissipation mechanisms is expressed by an effective damping force proportional to the drift velocity of the charge current. According to this picture, the equation of motion for the electric polarization **P** (electric dipole moment per unit volume) can be derived from Newton's law:

$$\partial_t^2 \mathbf{P} + \nu \partial_t \mathbf{P} = \varepsilon_0 \omega_p^2 \mathbf{E}, \quad (1)$$

Here $\nu$ is the damping coefficient, $\omega_p$ is the plasma frequency, $\varepsilon_0$ is the dielectric constant of the empty space, and $\mathbf{E}$ is the electric field. The permittivity

$$\varepsilon(\omega) = \varepsilon_0 \left(1 - \frac{\omega_p^2}{\omega^2 + i\nu\omega}\right) \quad (2)$$

of Drude model can be derived from Eq.(1) for harmonic fields. We now construct the Lagrangian density and dissipation function density for deriving Eq. (1) and all the Maxwell equations involving the $\mathbf{E}$, $\mathbf{B}$, $\mathbf{D}$, and $\mathbf{H}$ fields. Note that for the present case no magnetization $\mathbf{M}$ need to be considered so the simple constitutive relations $\mathbf{D} = \varepsilon_0 \mathbf{E} + \mathbf{P}$ and $\mathbf{B} = \mu_0 \mathbf{H}$ hold. In addition, since the Gauss' law $\nabla \cdot \mathbf{B} = 0$ for the $\mathbf{B}$ field and the Faraday's induction law $\nabla \times \mathbf{E} + \partial_t \mathbf{B} = 0$ hold automatically by assuming the defining relations $\mathbf{B} = \nabla \times \mathbf{A}$ and $\mathbf{E} = -\nabla \varphi - \partial_t \mathbf{A}$, where $\varphi$ and $\mathbf{A}$ are the scalar and vector potential respectively, here we need only to derive the Gauss' law and the Ampere's law

$$\nabla \cdot \mathbf{D} = 0 \quad (3)$$
$$\nabla \times \mathbf{H} = \partial_t \mathbf{D} \quad (4)$$

for the $\mathbf{D}$ and $\mathbf{H}$ fields from the Lagrange equations associated with the dissipation function $\mathcal{F}$:

$$\partial_t \left(\frac{\partial \mathcal{L}}{\partial(\partial_t \psi_\alpha)}\right) + \partial_j \left(\frac{\partial \mathcal{L}}{\partial(\partial_j \psi_\alpha)}\right) - \frac{\partial \mathcal{L}}{\partial \psi_\alpha} = -\frac{\partial \mathcal{F}}{\partial(\partial_t \psi_\alpha)} \quad (5)$$

Here $\psi_\alpha = [\varphi, \mathbf{A}, \mathbf{P}]$ are the dynamical field variables involved in the Lagrangian density $\mathcal{L}$, and $\dot{\psi}_\alpha = [\dot{\varphi}, \dot{\mathbf{A}}, \dot{\mathbf{P}}]$ appear in the dissipation function density $\mathcal{F}$. It is obvious that the Lagrangian density should contain the term $\mathcal{L}_0 = \frac{\varepsilon_0}{2} E^2 - \frac{\mu_0}{2} H^2$ for the electromagnetic fields themselves, together with the interaction energy term $\mathcal{L}_{int} = \mathbf{P} \cdot \mathbf{E}$ for the dipole-field interaction, and the kinetic energy term of the form $\mathcal{L}_k = \eta(\partial_t \mathbf{P})^2$ for the moving electrons, here $\eta$ is a constant to be determined. The dissipation function density must be of the form $\mathcal{F} = \sigma(\partial_t \mathbf{P})^2$ because a damping force proportional to the current or velocity leads to a power loss proportional to the square of the velocity similar to the power of Joule heat. Using the 'trial-and-error' trick we find $\eta = \frac{1}{2\omega_p^2 \varepsilon_0}$ and $\sigma = \frac{\nu}{2\omega_p^2 \varepsilon_0}$, and the Lagrangian and dissipation function density in terms the $\psi_\alpha$ variables are:

$$\mathcal{L} = \frac{\varepsilon_0 (\nabla \varphi + \partial_t \mathbf{A})^2}{2} - \frac{(\nabla \times \mathbf{A})^2}{2\mu_0} + \frac{(\partial_t \mathbf{P})^2}{2\omega_p^2 \varepsilon_0} - \mathbf{P} \cdot (\nabla \varphi + \partial_t \mathbf{A}) \quad (6)$$

$$\mathcal{F} = \frac{\nu}{2\omega_p^2 \varepsilon_0} (\partial_t \mathbf{P})^2 \quad (7)$$

Note that the dissipation function density is one half of the power loss we derived before. This feature will also hold in the following examples of dispersive-absorptive metamaterials as can be checked later. Substituting Eq. (6) and Eq. (7) into Eq.(5), we get all the dynamical equations for the polarization $\mathbf{P}$, electric displacement $\mathbf{D}$, and the magnetic fields $\mathbf{H}$ as in Eq.(1), Eq.(3) and Eq.(4).

The Hamiltonian density $\mathcal{H}$ can be obtained using the Legendre transformation:

$$\mathcal{H} = \pi_\alpha \dot{\psi}_\alpha - \mathcal{L}$$
$$= \frac{\varepsilon_0 \mathbf{E}^2}{2} + \frac{\mu_0 \mathbf{H}^2}{2} + \frac{(\partial_t \mathbf{P})^2}{2\omega_p^2 \varepsilon_0} + \nabla \varphi \cdot \mathbf{D} \quad (8)$$

In Eq.(8), the canonical momenta $\pi_\alpha$ conjugate to the $\psi_\alpha$ fields are $\pi_\alpha = \left[0, -\mathbf{D}, \partial_t \mathbf{P}/\omega_p^2 \varepsilon_0\right]$. Note that the canonical momentum $\pi_\varphi$ conjugates to $\varphi$ is lacking. This reflects the fact that $\varphi$ plays the role of a Lagrangian multiplier which can take arbitrary value. Since the constraint $\nabla \cdot \mathbf{D} = 0$ (Eq.(3)) for the displacement field leads to $\nabla \varphi \cdot \mathbf{D} = \nabla \cdot (\varphi \mathbf{D}) - \varphi \nabla \cdot \mathbf{D} = \nabla \cdot (\varphi \mathbf{D})$, and according to Lagrangian field theory a divergence term like $\nabla \cdot (\varphi \mathbf{D})$ has no effect on the equations of motion, we can drop this term in Eq.(8). The resultant Hamiltonian density is obviously an energy density including the term $\frac{\varepsilon_0 \mathbf{E}^2}{2} + \frac{\mu_0 \mathbf{H}^2}{2}$ for the energy density of the electromagnetic fields themselves, and the kinetic energy density $(\partial_t \mathbf{P})^2 / 2\omega_p^2 \varepsilon_0$ of the plasma electrons.

## 3. WIRE-SRR METAMATERIAL

We now construct the Lagrangian density and dissipation function for the metamaterial consisting of wires and split-ring resonators (SRRs). The wire structure alone as an effective medium in the long-wavelength limit has the effective permittivity of Eq. (2), but the effective plasma frequency $\omega_p$ and the damping coefficient $\nu$ are now dependent on some geometrical factors and the material characters of the wires. The effective medium of the split-ring resonators (SRRs) array, on the other hand, has the permeability

$$\mu(\omega) = \mu_0 \left(1 - \frac{F\omega^2}{\omega^2 - \omega_0^2 + i\gamma\omega}\right), \quad (9)$$

where $F$ is the filling fraction, $\omega_0$ is the resonance frequency, and $\gamma$ is the effective damping coefficient of the SRRs. The equation of motion for the magnetization field $\mathbf{M} = \mathbf{B}/\mu_0 - \mathbf{H}$ is

$$\partial_t \mathbf{M} + \gamma \mathbf{M} + \omega_0^2 \int \mathbf{M} dt = -F \partial_t \mathbf{H}. \quad (10)$$

which yields the permeability in Eq. (9) for the harmonic fields. Note that the magnetization $\mathbf{M}$ is proportional to the charge current and the cross section area of each SRR, therefore the first order differential equation Eq.(10) also represents the second order differential equation for the charge stored in the capacitance of the SRR. We define a new field

$$\mathbf{Q} = \mu_0 \int \mathbf{M} dt \quad (11)$$

to represent the dynamic field proportional to the stored charge in the SRR. It is obvious that the SRR structure contributes a partial Lagrangian of the form $\mathcal{L}_{SRR} = \alpha \mathbf{M}^2 - \beta \mathbf{Q}^2$. The $\mathbf{M}^2$ term representing the 'kinetic energy' of the moving charges circulating around the SRR is in fact the magnetic field energy associated with inductance of the SRR, while the 'potential energy' term proportional to $\mathbf{Q}^2$ is the electric field energy stored in the capacitance of the SRR. In addition, the wires array contributes partial Lagrangian density $\mathcal{L}_{wire} = \frac{1}{2\omega_p^2 \varepsilon_0}(\partial_t \mathbf{P})^2 + \mathbf{P} \cdot \mathbf{E}$, which includes the 'kinetic energy' $\frac{1}{2\omega_p^2 \varepsilon_0}(\partial_t \mathbf{P})^2$ from the magnetic field energy associated with the wire inductance, and the polarization energy $-\mathbf{P} \cdot \mathbf{E}$ due to the work done by the electric field to the moving charges along the wires (i.e, the 'qV' energy).

The remaining terms in the complete Lagrangian density must include the contributions from the 'pure electromagnetic fields', and the interaction energy between the fields and the SRR array, so it should be of the form $\mathcal{L}_{\text{EM\_SRR}} = \frac{\varepsilon_0}{2}\mathbf{E}^2 - \frac{1}{2\mu_0}\mathbf{B}^2 + \lambda \mathbf{B} \cdot \mathbf{M}$. The defining relations $\mathbf{E} = -\nabla \varphi - \partial_t \mathbf{A}$ and $\mathbf{B} = \nabla \times \mathbf{A}$ still hold, but the magnetic induction $\mathbf{B}$ now contains the magnetization $\mathbf{M}$ (i.e., $\mathbf{B} = \mu_0(\mathbf{H} + \mathbf{M})$), which is lacking in the Drude model. There is the ambiguity about is it the $\mathbf{B}$ field or $\mathbf{H}$ field appearing in this partial Lagrangian. The answer is irrelevant because the difference between the two choices is a term proportional to $\mathbf{M}^2$, and a term of this kind can always be absorbed into the $\mathcal{L}_{\text{SRR}}$ partial Lagrangian by simply tuning the coefficient $\alpha$.

Now the complete Lagrangian density can be determined by requiring that all the dynamic equations for the fields (Eq.(3) and (4)) and the matters (Eq.(1) and (10)) should be obtained form the Lagrangian equations (Eq.(5)) that corresponds to the dynamical field variables $\psi_\alpha = [\varphi, \mathbf{A}, \mathbf{P}, \mathbf{Q}]$ if the dissipation function density is also provided. Since the dissipation in the Wire-SRR system is due to the Joule heat power generated in the wires and SRRs, it is clear that the dissipation function density should contain the terms proportional to $(\partial_t \mathbf{P})^2$ and $\mathbf{M}^2$, respectively. Based on these considerations, the complete Lagrangian density is

$$\mathcal{L} = \frac{\varepsilon_0 \mathbf{E}^2}{2} - \frac{\mu_0 \mathbf{H}^2}{2} + \frac{(\partial_t \mathbf{P})^2}{2\omega_p^2 \varepsilon_0} + \mathbf{P} \cdot \mathbf{E} + \frac{\mu_0 \mathbf{M}^2}{2F} - \frac{\omega_0^2 \mathbf{Q}^2}{2\mu_0 F}$$

$$= \frac{\varepsilon_0 (\nabla \varphi + \partial_t \mathbf{A})^2}{2} - \frac{(\nabla \times \mathbf{A} - \partial_t \mathbf{Q})^2}{2\mu_0} \quad (12)$$

$$+ \frac{(\partial_t \mathbf{P})^2}{2\omega_p^2 \varepsilon_0} - \mathbf{P} \cdot (\nabla \varphi + \partial_t \mathbf{A}) + \frac{[(\partial_t \mathbf{Q})^2 - \omega_0^2 \mathbf{Q}^2]}{2\mu_0 F}$$

and the dissipation function density is

$$\mathcal{F} = \frac{\nu}{2\omega_p^2 \varepsilon_o}(\partial_t \mathbf{P})^2 + \frac{\gamma \mu_0}{2F}\mathbf{M}^2 \quad (13)$$

It is very interesting to note that the field-matter coupling term of the form $\mathbf{H} \cdot \mathbf{M}$ hidden in the Lagrangian density if we use $\mathbf{H}$ instead of $\mathbf{B}$ to express the Lagrangian density, as is indicated in the first line of Eq.(12). In addition, the dissipation function density is still one half of the power loss of the system, just like we have already learned in the Drude model problem.

Now the canonical momenta conjugate to $\psi_\alpha = [\varphi, \mathbf{A}, \mathbf{P}, \mathbf{Q}]$ are

$\pi_\alpha = \left[0, -\mathbf{D}, \partial_t \mathbf{P}/\omega_p^2 \varepsilon_0, \mathbf{H} + \frac{\mathbf{M}}{F}\right]$. The irrelevant $\pi_\varphi$ is still lacking as before. We also noticed that $\pi_\mathbf{Q}$ is equal to the magnetic field inside an SRR. Applying Legendre transformation again, we get the Hamiltonian density

$$\mathcal{H} = \frac{\varepsilon_0 \mathbf{E}^2}{2} + \frac{(\partial_t \mathbf{P})^2}{2\omega_p^2 \varepsilon_0} + \frac{\mu_0 \mathbf{H}^2}{2} + \frac{\mu_0}{2}\mathbf{H} \cdot \mathbf{M}$$

$$+ \frac{\mu_0 \mathbf{M}^2}{2F} + \frac{\mu_0 \omega_0^2 (\int \mathbf{M} dt)^2}{2F} + \nabla \varphi \cdot \mathbf{D} \quad (14)$$

Except for the extra term $\nabla \varphi \cdot \mathbf{D}$ that can be dropped, this result is exactly the same as the energy density we have obtained previously (The sum of the Eq.(13) and Eq.(15) in Ref.[14]). After dropping this term the Hamiltonian density becomes identical to the energy density we obtained before.

## 4. SINGLE-RESONANCE CHIRAL METAMATERIAL

This metamaterial is composed of an array of uncoupled helix resonators. Each resonator is a finite-length helix of conducting wire, which can be viewed as a twisted SRR. This feature implies that each helix resonator is a RLC circuit equipped with inductance, capacitance, and resistance, and both electric field and changing magnetic field can induce current as well as electric dipole in it. Accordingly, the magnetization $\mathbf{M}$ in this medium must be proportional to the changing rate of the effective polarization $\partial_t \mathbf{P}$. The constitutive relations

$$\mathbf{D}_\omega = \varepsilon \mathbf{E}_\omega + i(\kappa/c)\mathbf{H}_\omega, \quad (15)$$

$$\mathbf{B}_\omega = \mu \mathbf{H}_\omega - i(\kappa/c)\mathbf{E}_\omega. \quad (16)$$

for harmonic electromagnetic waves are given by

$$\varepsilon = \varepsilon_0 \left(1 - \frac{\omega_p^2}{\omega^2 - \omega_0^2 + i\Gamma\omega}\right), \quad (17)$$

$$\mu = \mu_0 \left(1 - \frac{F\omega^2}{\omega^2 - \omega_0^2 + i\Gamma\omega}\right), \quad (18)$$

$$\kappa = \frac{A\omega}{\omega^2 - \omega_0^2 + i\Gamma\omega}, \quad (19)$$

where $\omega_p$ and $\omega_0$ are two characteristic frequencies, $F$ is the filling fraction of the resonators, $\Gamma$ is a dissipation coefficient, and $A = \pm\sqrt{F}\omega_p$ relates $\mathbf{M}$ and $\partial_t \mathbf{P}$ or $\mathbf{Q}$ (ref. Eq.(11)) and $\mathbf{P}$

$$\partial_t \mathbf{P} = -\frac{\omega_p^2}{Ac}\mathbf{M}, \quad \mathbf{P} = -\frac{\omega_p^2}{A\mu_0 c}\mathbf{Q} \quad (20)$$

The constitutive relations Eq. (15) to Eq.(19) are derived from the dynamical equations

$$\partial_t^2 \mathbf{P} + \Gamma \partial_t \mathbf{P} + \omega_0^2 \mathbf{P} = \varepsilon_0 \omega_p^2 \mathbf{E} + \frac{A}{c}\partial_t \mathbf{H}, \quad (21)$$

$$\partial_t \mathbf{M} + \Gamma \mathbf{M} + \omega_0^2 \int \mathbf{M} dt = -F\partial_t \mathbf{H} - \frac{A}{\mu_0 c}\mathbf{E}. \quad (22)$$

corresponding to the two fundamental relations $\mathbf{D} = \varepsilon_0 \mathbf{E} + \mathbf{P}$ and $\mathbf{B} = \mu_0(\mathbf{H} + \mathbf{M})$. Furthermore, Eq.(21) and (22) are in fact one and the same thing derived from the circuit equation of the helix resonator, as that indicated by the constraint Eq.(20). This fact leads to an equality $(\partial_t \mathbf{P})^2/\omega_p^2 \varepsilon_0 = \mu_0 \mathbf{M}^2/F$, which implies that any term proportional to $(\partial_t \mathbf{P})^2$ in the Lagrangian or the dissipation function is not different from a term proportional to $\mathbf{M}^2$. Therefore, only one of these two terms will show up in the Lagrangian and the dissipation function. In addition, Eq. (2) is a Drude-model type dispersion, while Eq.

(17) corresponds to a Lorentz-model dispersion of resonance frequency $\omega_0$, therefore a 'potential energy' term proportional to $\mathbf{P}^2$ or $\mathbf{Q}^2$ must be included in the Lagrangian density. Furthermore, the independent dynamical fields can either be chosen as $\psi_\alpha = [\varphi, \mathbf{A}, \mathbf{P}]$ or $\tilde{\psi}_\alpha = [\varphi, \mathbf{A}, \mathbf{Q}]$ because they are equivalent. The complete Lagrangian density and dissipation function are given by

$$\mathcal{L} = \frac{\varepsilon_0 \mathbf{E}^2}{2} - \frac{\mu_0 \mathbf{H}^2}{2} + \mathbf{P} \cdot \mathbf{E} + \frac{\mu_0 \mathbf{M}^2}{2F} - \frac{\omega_0^2 \mathbf{Q}^2}{2\mu_0 F}$$

$$= \frac{\varepsilon_0 (\nabla \varphi + \partial_t \mathbf{A})^2}{2} - \frac{(\nabla \times \mathbf{A} - \partial_t \mathbf{Q})^2}{2\mu_0}$$

$$- \mathbf{P} \cdot (\nabla \varphi + \partial_t \mathbf{A}) + \frac{\left[(\partial_t \mathbf{Q})^2 - \omega_0^2 \mathbf{Q}^2\right]}{2\mu_0 F}$$

(23)

$$\mathcal{F} = \frac{\mu_0 \Gamma}{2F} \mathbf{M}^2 ,$$

(24)

as can be examined by substituting them into Eq.(5) and check the equations of motion.

Choosing $\tilde{\psi}_\alpha = [\varphi, \mathbf{A}, \mathbf{Q}]$ as the dynamical fields, the canonical momenta conjugate to them are found to be the same as that in the wire-SRR metamaterial, i.e., $\pi_\alpha = \left[0, -\mathbf{D}, \mathbf{H} + \frac{\mathbf{M}}{F}\right]$. The Legendre transformation, leads to the Hamiltonian density

$$\mathcal{H} = \frac{\varepsilon_0 \mathbf{E}^2}{2} + \frac{\mu_0 \mathbf{H}^2}{2} + \frac{\mu_0}{2} \mathbf{H} \cdot \mathbf{M}$$

$$+ \frac{\mu_0 \mathbf{M}^2}{2F} + \frac{\mu_0 \omega_0^2 \left(\int \mathbf{M} dt\right)^2}{2F} + \nabla \varphi \cdot \mathbf{D}$$

(25)

which is still the same as our previous result of the energy density (Eq.(17) of Ref. [15]) if the redundant divergence term $\nabla \varphi \cdot \mathbf{D} = \nabla \cdot (\varphi \mathbf{D})$ is dropped.

## 5. CONCLUSION

In this paper, we proposed the Lagrangian density and the associated dissipation function for two typical dispersive-absorptive metamaterials: the wire-SRR metamaterial and the single-resonance chiral metamaterial. Applying Legendre transformation to these two cases, the resultant Hamiltonians are identical to the energy densities we obtained previously based on Poynting theorem and the knowledge of the dissipation. This work not only provides an systematic method for deriving the energy density of the electromagnetic fields in dispersive-absorptive metamaterials, but also demonstrates how to construct the low energy effective Hamiltonian for the electrodynamics systems of metamaterials for the further research of the quasiparticles or elementary excitations in the dispersive metamaterials.

**Funding Information.**
National Science Council (NSC) of Taiwan: 103-2918-I-008-012.

**Acknowledgment**.
I would like to thank my colleague C. S. Tang for some useful discussions and insightful comments.